\documentclass[amssymb,prb,twocolumn,floats]{revtex4}
\usepackage{bm}
\usepackage{graphicx}

\begin{document}
\preprint{March 20, 2004}

\title{Photoluminescence of p-doped quantum wells with strong spin splitting}

\author{P. Kossacki$^{1,2,3}$}
\email[]{Piotr.Kossacki@fuw.edu.pl}

\author{H. Boukari$^{1}$ }
\author{M. Bertolini$^{1}$ }
\author{D. Ferrand$^{1}$ }
\author{J. Cibert$^{1,4}$ }
\author{S. Tatarenko$^{1}$ }
\author{J.A. Gaj$^{2}$ }
\author{B.Deveaud$^{3}$ }
\author{V.Ciulin$^{3}$ }
\author{M. Potemski$^{5}$ }

\affiliation{
$^{1}$  Groupe "Nanophysique et Semiconducteurs", CEA-CNRS-Universit\'e Joseph Fourier Grenoble, Laboratoire de Spectrom\'etrie Physique, BP87, F-38402 Saint Martin d'H\`eres cedex, France.\\
$^{2}$  Institute of Experimental Physics, Warsaw University, Ho\.za 69, PL-00-681 Warszawa, Poland.\\
$^{3}$  Institute of Quantum Electronics and Photonics, Ecole Polytechnique F\'ed\'erale de Lausanne (EPFL), CH-1015 Lausanne, Switzerland.\\
$^{4}$  Laboratoire Louis N\'eel, CNRS, BP166,  F-38042 Grenoble Cedex 9, France. \\
$^{5}$  Grenoble High Magnetic Field Laboratory, MPI-FKF/CNRS, 25
Avenue des Martyrs, F-38042 Grenoble Cedex 9, France. }

\date{\today}

\begin{abstract}

The spectroscopic  properties of a spin polarized two-dimensional
hole gas are studied in modulation doped Cd$_{1-x}$Mn$_x$Te
quantum wells with variable carrier density up to
$5~10^{11}$~cm$^{-2}$. The giant Zeeman effect which is
characteristic of diluted magnetic semiconductors, induces a
significant spin splitting even at very small values of the
applied field. Several methods of measuring the carrier density
(Hall effect, filling factors of the Landau levels at high field,
various manifestations of Moss-Burstein shifts) are described and
calibrated. The value of the spin splitting needed to fully
polarize the hole gas, evidences a strong enhancement of the spin
susceptibility of the hole gas due to carrier-carrier interaction.
At small values of the spin splitting, whatever the carrier
density (non zero) is, photoluminescence lines are due to the
formation of charged excitons in the singlet state. Spectral
shifts in photoluminescence and in transmission (including an
"excitonic Moss-Bustein shift") are observed and discussed in
terms of excitations of the partially or fully polarized hole gas.
At large spin splitting, and without changing the carrier density,
the singlet state of the charged exciton is destabilized in favour
of a triplet state configuration of holes. The binding energy of
the singlet state is thus measured and found to be independent of
the carrier density (in contrast with the splitting between the
charged exciton and the neutral exciton lines). The state stable
at large spin splitting is close to the neutral exciton at low
carrier density, and close to an uncorrelated electron-hole pair
at the largest values of the carrier density achieved. The triplet
state gives rise to a characteristic double-line structure with an
indirect transition to the ground state (with a strong phonon
replica) and a direct transition to an excited state of the hole
gas.

\end{abstract}


\draft





\maketitle

\section{Introduction.}

The spectroscopy of two dimensional systems with carriers has been
studied intensively during past years. At low carrier density,
optical spectra are dominated by a line related to the charged
exciton (trion) transition \cite{Kheng93}. In the limit of very
low carrier density, the charged exciton is a three particle
complex involving a pre-existing carrier and the photocreated
electron-hole pair, the two identical particles being arranged in
a singlet configuration. Some trion properties are well
understood. For example, many papers were devoted to theoretical
and experimental studies on the binding energy under different
conditions and in different materials
\cite{Astakhov02b,Redlinski01,Stebe98,Esser00,Riva00}. Detailed
investigations were also performed on the selection rules
\cite{Lovisa97, Kheng95,Astakhov02a}, and on the dynamics
\cite{Finkelstein98,Vanelle00,Yoon96,Ciulin2000}, in particular
the formation time \cite{Kutrowski2002,Jeukens2002,Kossacki2002}
and spin relaxation \cite{Vanelle00,Ciulin2002}. Most of those
properties are well established and can be used for identification
purposes.

In high  magnetic field, a triplet state has been described
\cite{Shields95a,Sanvitto2002,Crooker00} and its stability with
respect to the singlet state has been discussed in terms of
orbital wave functions.

The range of higher carrier densities is less understood. As the
carrier density increases, an increase of the energy splitting
between the absorption lines of the charged and neutral exciton
has been reported in various materials
\cite{Cole98,Huard2000,Kossacki99,Astakhov02b}. This effect does
not necessarily mean that the binding energy changes: it is
important to analyze both the initial and final states involved in
the spectroscopic transitions as well as the change of lineshapes
versus carrier density. Thus, recoil gives rise to low energy
tails while exciton-electron unbound states cause a strong high
energy tail which adds a large contribution to the oscillator
strength of the neutral exciton
\cite{Suris2001,Esser2001,Cox2004}. At even higher densities, the
neutral exciton lines disappears and the charged exciton line
progressively transforms into the Fermi edge singularity which
marks the onset of the electron-hole continuum
\cite{Kossacki99,Huard2000b,Wojtowicz98}. All these effects point
to a non-negligible interaction of the charged exciton with the
carrier gas.

We wish to discuss here selected problems related to the
photoluminescence (PL) of doped quantum wells (QW), thanks to new
experimental results associated to a strong spin splitting in
magnetic fields small enough to prevent Landau quantization. This
is obtained thanks to the giant Zeeman effect in
Cd$_{1-x}$Mn$_{x}$Te QWs which are modulation doped p-type. The
carrier density was adjusted through the design of the structure
using either nitrogen acceptors or surface states, and in some
samples it was controlled either optically or through a bias
applied across a pin diode. A thorough calibration of the carrier
density was done.

We show that for small values of the spin splitting ($<$~3~meV),
the PL line in both circular polarizations is due to the singlet
state of the charged exciton. However, a shift appears between the
absorption and PL lines \cite{Kossacki99, Kochereshko2000}, and
increases with the carrier density quite similarly to the
Moss-Burstein shift of band-to-band transitions. This shift is
discussed in terms of excitations in the final states of the
transitions. It vanishes in $\sigma^-$ polarization when the spin
splitting is large enough that the hole gas be completely
polarized: the relevant value of the spin splitting is discussed
with respect to the Fermi energy. Then, above a certain value of
the spin splitting, the lowest initial state in the PL transition
is no more the singlet state of the charged exciton, but a state
where all holes are in the majority spin subband. That means that
the singlet arrangement of the two holes involved in the charged
exciton is replaced by a parallel arrangement of their spins
(triplet state). Contrary to the case of nonmagnetic quantum
wells, for which the triplet state of the charged exciton is
stabilized by orbital effects at very high magnetic field
\cite{Sanvitto2002,Homburg2000,Riva2001}, in the present case the
singlet state is destabilized by spin effects. This gives rise to
several features which agree with a mechanism involving
band-to-band transitions (double line, with the lower component
having the ground state of the hole gas as its final state, and
the upper one leaving the carrier gas in an excited state; and
existence of a Moss-Burstein shift).  Finally, different energy
scales involved are discussed with reference to the Fermi energy.

This paper is organized as follows. Section II briefly describes
the samples and the experimental set-up. Section III is devoted to
an advanced characterization of the samples and of the mechanisms
involved in the spectroscopic properties, based on our previous
knowledge of p-doped Cd$_{1-x}$Mn$_{x}$Te QWs: determination of
the diamagnetic shift and of the normal Zeeman effect, measure of
the giant Zeeman effect and derivation of the exact density of
free spins, behavior at large values of the applied magnetic field
and derivation of the position of the Landau levels and of the
carrier density from the filling factors, and finally
determination of the carrier density from Hall effect. Section IV
is the core section of this paper: it describes the PL spectra
which characterize a Cd$_{1-x}$Mn$_{x}$Te QW in the presence of
holes, first in the low density limit where the spectra can be
understood from the competition between neutral and charged
excitons, then at higher carrier density where new features are
observed and analyzed in terms of the initial and final states
involved in the PL transition, and used to further characterize
the hole gas (density and polarization). Some consequences of the
observation of these new features are discussed in section V.

\section{Samples and experimental set up.}
Samples have been grown by molecular-beam epitaxy from CdTe, ZnTe,
Te, Cd, Mg and Mn sources, on (001)-oriented Cd$_{1-z}$Zn$_{z}$Te
substrates, most of them with $z=0.12$, with a buffer layer of
similar composition and (Cd,Zn,Mg)Te barriers. Characteristics of
selected samples are summarized in table I. All samples were grown
pseudomorphically, therefore the lattice constant of the substrate
governs the uniaxial strain in the single QW: due to confinement
and strain, the heavy-hole state is the ground state in the
valence band, the light-hole band being 30 to 40~meV higher in
energy. Most samples were modulation doped with nitrogen in the
barriers \cite{Arnoult99}. In others, the hole gas was provided by
proper adjustment of the (Cd,Mg)Te cap layer thickness, with
surface states playing the role of acceptors; our previous studies
\cite{Maslana03} allowed us to optimize the structure of these
samples in order to obtain the largest hole density, by placing
the QW 25 nm below the surface and 200 nm away from the (Cd,Zn)Te
buffer layer. In both cases, illumination with light of energy
larger than the energy gap of the barrier material results in the
depletion of the hole gas \cite{Shields95b, Shields96,
Kossacki99}: This method with a suitable calibration
\cite{Kossacki99} was used in order to tune the carrier density in
the QW from the maximum value (a few 10$^{11}$ cm$^{-2}$) down to
the low 10$^{10}$ cm$^{-2}$ range. The second way to control the
carrier concentration was to insert the QW in a p-i-n diode. In
such a structure \cite{Boukari02}, the back barrier doped with
aluminum (n-type) was 320 nm away from the QW, and the spacer
between the QW and the p-doped layer was reduced to 10~nm. A 10~nm
semi-transparent gold film was evaporated on top of the p-i-n
diodes, and then 2$\times$2 mm2 squares were formed by Ar-ion
etching down to the n-type layer, a procedure followed by the
deposition of In contacts. In these diodes, non-linear
current-voltage characteristics were observed up to room
temperature, and applying a small bias resulted in variations of
the carrier density from below 10$^{10}$ cm$^{-2}$ up to
4$\times$10$^{11}$ cm$^{-2}$. In all samples the hole densities
are such that the carriers occupy only the ground-state heavy-hole
subband.

All properties discussed below were determined or checked by
magneto-optical spectroscopy performed in the Faraday
configuration (magnetic field and light propagation perpendicular
to the sample surface). The samples were mounted strain-free in a
superconducting magnet and immersed in liquid helium for low
temperature measurements. The experimental setup allowed us to
perform typical cw measurements such as reflectivity, PL and PL
excitation (PLE). The Cd$_{0.88}$Zn$_{0.12}$Te substrates are
transparent at the relevant wavelengths so that in most cases we
could also perform transmission experiments. In PL, the optical
excitation was achieved with a tunable Ti-sapphire laser
providing about 2~mW cm$^{-2}$, or a HeNe laser with similar
intensity. A halogen lamp (filtered to avoid carrier depletion,
or not in order to induce a depletion) was used as a source of
light for reflectivity and transmission measurements. Time
resolved PL was excited by a picosecond tunable Ti-sapphire laser
with a 2 ps pulse width, a repetition rate of 80~MHz, and an
averaged power density less than 100 mW cm$^{-2}$. The signal was
collected through a spectrometer by a 2D streak camera with 10 ps
resolution. The measurements in high magnetic fields were
performed in the Grenoble High Field Laboratory using a 20 T
Bitter coil.

\section{Basic spectroscopic properties and characterization of the 2D hole gas.}
In this section we summarize the characterization of our samples
and of some of their spectroscopic properties, based on the
previous knowledge of undoped and p-doped CdTe and
Cd$_{1-x}$Mn$_{x}$Te QWs. We first determine the parameters
describing the properties of the excitons under the direct
influence of an applied magnetic field (diamagnetic shift, Lande
factor) in a non-magnetic CdTe QW. Then we show that exploiting
the giant Zeeman effect we achieve a good description of the
excitons in a diluted magnetic QW at low hole density. Finally we
turn to a thorough determination of the hole density, using
methods which do not depend on materials parameters, such as
plotting the positions of integer filling factors of Landau
levels, or deriving the density from the Hall resistance.

\subsection{Excitonic regime: normal Zeeman effect and diamagnetic shift.}

Fig. 1a shows typical transmission spectra measured at different
values of the magnetic field  on sample N0, which is a 8 nm thick
CdTe QW, moderately doped p-type (nitrogen doping on both sides
with 50 nm spacer layers). The whole structure (QW,
Cd$_{0.69}$Zn$_{0.08}$Mg$_{0.23}$Te barriers,
Cd$_{0.88}$Zn$_{0.12}$Te buffer layer) is coherently strained on
the Cd$_{0.88}$Zn$_{0.12}$Te substrate, so that the
light-hole~/~heavy-hole splitting is about 40 meV. At zero field,
two lines are observed. As confirmed below, the higher-energy
line is related to the neutral exciton and the lower-energy one
to a charged exciton. Under magnetic field, the neutral exciton
is observed in both polarizations, while the low energy line
progressively disappears in $\sigma^+$ polarization. This
behavior is opposite to that of the negatively charged exciton
X$^{-}$, so that it was attributed to the positively charged
exciton in reference \cite{Haury98}, as confirmed and discussed
quantitatively in the following.

\begin{figure}[ht]
\includegraphics[width=85mm]{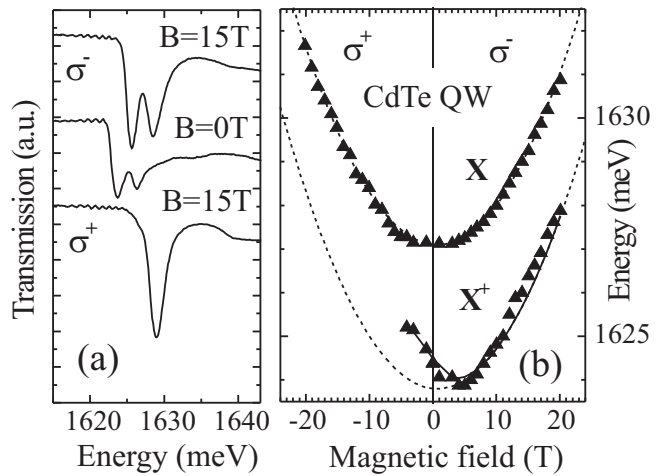}
\caption[width=85mm]{(a) Typical transmission spectra and (b)
magnetic field dependence of the transition energies, observed at
1.9K on sample N0, an 8 nm broad CdTe QW with
Cd$_{0.69}$Zn$_{0.08}$Mg$_{0.23}$Te barriers, moderately doped
p-type (two symmetrical N-doped layers at 50 nm) . In (b), the
right part (positive fields) displays the $\sigma^+$ polarization
data and the left part (negative fields) the $\sigma^-$ ones.
Dashed lines are quadratic fits; the solid line through the
charged exciton positions is obtained when taking into account the
dependence of the X~/~X$^{+}$ splitting on the hole polarization.}
\label{fig01}
\end{figure}

Fig. 1b shows the magnetic field dependence of the transition
energies. By convention, we attribute positive field range to the
$\sigma^-$ polarization. The neutral exciton energy follows a
quadratic dependence:

$$
\emph{E}_{\sigma\pm} = E_{X} \mp a_{z} B + a_{dia} B^{2}
$$

with $E_{X}$=1627~meV in the present sample, $a_{z}$=0.015~meV
T$^{-1}$, and $a_{dia}$=0.01057~meV T$^{-2}$. This value of the
diamagnetic shift is reasonable: it corresponds to an effective
radius of 6~nm, which is of the order of the Bohr radius. Note the
small value of the Zeeman shift (the linear term): defining the
excitonic Land\'e factor as
$$
g_{X} = \frac{E(\sigma^{+}) - E(\sigma^{-})}{\mu_{\mathrm{B}}B}
$$
we have $g_{X}$=0.5 , definitely smaller than in CdTe QWs of
similar width but with no strain \cite{Sirenko97} or a small one
\cite{Zhao96}. Indeed, in unstrained QWs the excitonic Land\'e
factor $g_{X}$=-1.5  mostly corresponds to the Land\'e factor of
the electron $g_{e}$=-1.4  (to be compared to $g_{e}$=-1.6 in bulk
CdTe), indicating an almost vanishing Zeeman splitting in the
valence band. In the present samples, coherently strained on the
Cd$_{0.88}$Zn$_{0.12}$Te substrate, using the same value $g_{e}$ =
-1.4  and our experimental value $g_{X}$=0.5 , we obtain
$g_{hh}=g_{X} + g_{e}$=-0.9. We use here the spin Land\'e factor
of the holes, i.e. the splitting between the two components of the
heavy holes is $g_{hh} \mu_{B} B$. This finite Zeeman splitting of
heavy holes is consistent with the magnetic circular dichroism of
the trion, opposite to that of the negatively charged exciton in
n-type CdTe QWs. Indeed, in similar samples with a low hole
density (in the 10$^{10}$ cm$^{-2}$ range) showing well defined
excitonic lines, we could directly estimate $g_{hh}$~=~-1.2 from
the field and temperature dependence of the dichroism
 by assuming that the distribution of carriers on
the two spin states obeys the Maxwell-Boltzmann statistics and
that the trion intensity in $\sigma^-$ polarization is
proportional to the density of $|$+$\frac{3}{2}\rangle$ holes
\cite{Haury98}.

The transition energy of the charged exciton in $\sigma^-$
polarization above 4 T obeys the same dependence as the neutral
exciton (with a diamagnetic shift at most 5$\%$ larger), 3.4 meV
lower in energy. This value indicates a moderate but significant
density of carriers, in the 10$^{10}$~cm$^{-2}$ range, if we
compare to values measured in (Cd,Mn)Te QWs \cite{Kossacki99}. It
was demonstrated there that the X-X$^{+}$ splitting in absorption
contains a contribution proportional to the density of the hole
gas in one spin subband (the one promoting the trion formation).
We use this result in order to describe the behavior of the trion
in the present CdTe QW below 4~T and in $\sigma^+$ polarization,
where a shift to high energy is observed. At 3-4~T in $\sigma^+$
polarization, the binding energy tends to 2.0~meV, which is the
value expected at vanishingly small population of the holes with
the relevant spin. The solid line represents the fit with the
X-X$^{+}$ splitting equal to 2.0~meV plus a term proportional to
the hole gas density in the relevant spin subband. The spin
polarization was calculated using the Mawwell-Boltzmann
distribution between the Zeeman-split hole levels (with
$g_{hh}$=-1.2). We obtain a perfect agreement, which is an
additional support for our identification of the two lines.

The values of the Zeeman splitting and diamagnetic shifts
determined here will be used systematically in the following for
excitonic states in DMS QWs, where the additional giant Zeeman
effect is dominant.

\subsection{Giant Zeeman effect.}

\begin{figure}[h]
\includegraphics[width=81mm]{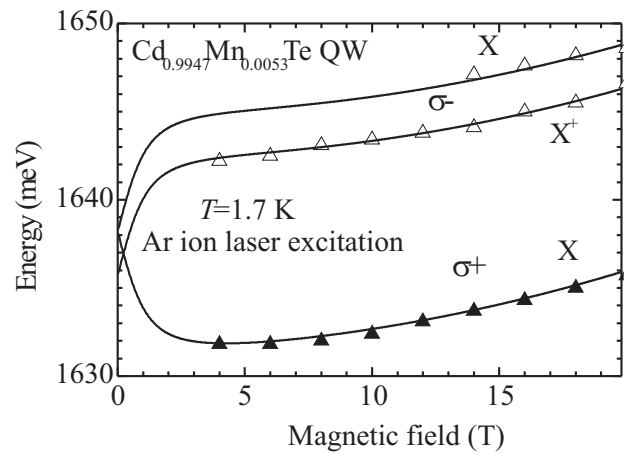}
\caption[]{Energy of the transmission line, as a function of
applied field, at 1.7~K, for sample N5, a p-doped, 8~nm wide,
Cd$_{1-x}$Mn$_{x}$Te QW. Ar-ion laser illumination strongly
reduces the carrier density to the low 10$^{10}$~cm$^{-2}$ range.
Solid lines are drawn with the same parameters as in Fig. 1 to
describe the normal Zeeman effect and diamagnetic shift, and a
Brillouin function with $x_{eff}$=0.0048 (and \emph{x}=0.0052,
$T_{0}$=0.17~K) to describe the giant Zeeman effect, and
adjustable values for the zero-field energies of X and X$^{+}$.}
\label{fig02}
\end{figure}

The parameters describing the giant Zeeman effect are easiest to
determine on spectra obtained with above-barrier illumination so
that the hole density is reduced. Fig.~\ref{fig02} is an example
of the position of the lines observed in transmission with
additional Ar-ion laser light shining on the surface of sample N5
(with 0.53$\%$ Mn in the QW). Due to the residual hole gas, we
observe the neutral exciton in $\sigma^+$ polarization and the
positively charged exciton in $\sigma^-$ polarization; the neutral
exciton is also observed at high field (much stronger than the
field for filling factor n = 1) in $\sigma^-$ polarization: this
agrees with previous observations at high field in the presence of
an electron gas \cite{Wojtowicz99, Kochereshko98}.

The field dependence of both the neutral and charged exciton
energies was fitted by adding the contribution from the giant
Zeeman effect to the normal Zeeman effect and diamagnetic shift,
with parameters from reference \cite{Gaj94}:

\begin{eqnarray*}
\emph{E}_{\sigma\pm} = E_{X} \mp a_{z} B + a_{dia} B^{2} \\
 \mp N_{0}(\alpha-\beta)\frac{1}{2}x_{eff}
\mathbf{B}_{\frac{5}{2}}(\frac{\frac{5}{2}g_{Mn}\mu_{\mathrm{B}}B}{k_{\mathrm{B}}(T+
T_{0})})
\end{eqnarray*}

where the parameters $a_{z}$ and $a_{dia}$ were determined above,
$\mathbf{B}_{\frac{5}{2}}$ denotes the Brillouin function,
$\emph{T}$ the temperature, $\emph{N}_{0}\alpha$=0.22 eV and
$\emph{N}_{0}\beta$=-0.88 eV describe the spin-carrier coupling in
the conduction and the valence band, respectively. The fitting
parameter x$_{eff}$ (density of free spins) and $\emph{T}_{0}$ are
related to the Mn content $\emph{x}$ by the following expressions,
calculated from \cite{Gaj94}:

\begin{eqnarray*}
x_{eff} = x[0.2635\exp(-43.34x) \\+ 0.729\exp(-6.190x) + 0.00721]
\end{eqnarray*}
and
$$
T_{0} = \frac{35.37x}{1 + 2.752x} ~ [\mathrm{K}]
$$

We did not include magnetization steps which are expected at high
field and very low temperature \cite{Shapira87,Zehnder96}, and
which would induce an additional linear increase of the
magnetization at the temperatures of interest in the present
study. The Mn content in our QWs remains below 1$\%$, so that the
amplitude of the steps remains small.

Mn contents quoted below are determined from the fit of the giant
Zeeman effect. They were found to agree with those expected from
growth conditions.

\subsection{High field Landau levels.}

\begin{figure}[ht]
\includegraphics[width=81mm]{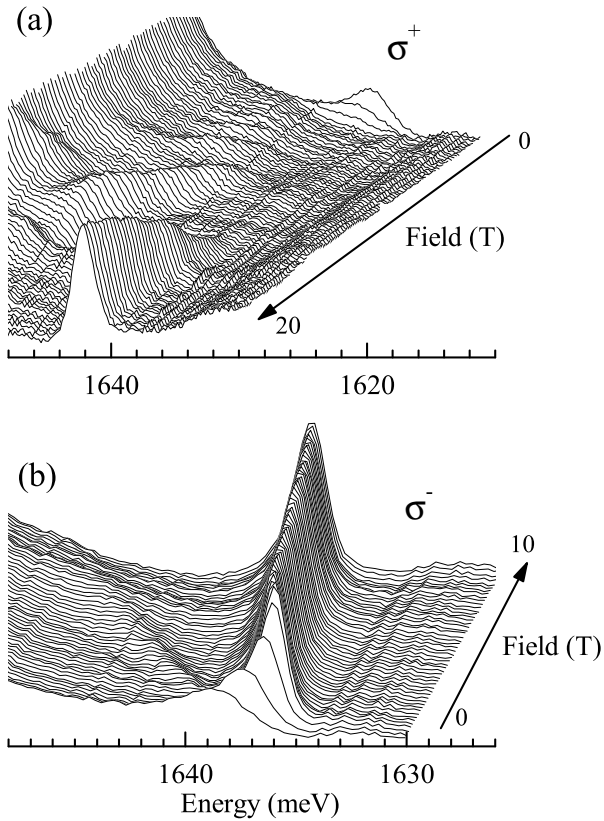}
\caption[]{(a) Transmission spectra , at 1.7 K and magnetic fields
from 0 to 20~T, in $\sigma^+$ polarization, for a
Cd$_{0.9982}$Mn$_{0.0018}$Te QW (sample N2); spectra have been
shifted in both directions to enhance the principal features; (b)
same as (a), in $\sigma^-$ polarization, from 0 to 10~T; shifts
are different in order to enhance the features of interest.}
\label{fig03}
\end{figure}

Fig.3 displays transmission spectra of sample N2, i.e., a
Cd$_{0.9982}$Mn$_{0.0018}$Te QW with
Cd$_{0.66}$Mg$_{0.27}$Zn$_{0.07}$Te barriers doped p-type on one
side with a 20 nm spacer. In these spectra, the smooth rise of the
background at high energy is due to absorption in the
Cd$_{0.88}$Zn$_{0.12}$Te substrate. At zero field, one observes an
asymmetric absorption feature, as expected in the presence of a
significant density of carriers. In this sample, in spite of the
low Mn content, the giant Zeeman effect is large enough that the
hole gas is fully polarized already at $\emph{B}$=0.6 T
\cite{Kossacki99}; as a result, circularly polarized spectra
differ qualitatively.

With applied field, the $\sigma^-$ spectra (Fig.3b) feature mostly
a single, intense line. The intensity of this line increases first
(up to 0.6 T) as the hole gas becomes more and more polarized, so
that the density of holes on the spin subband involved in the
transition ($|-\frac{3}{2}\rangle$ heavy holes) decreases to zero;
then the intensity stays practically constant up to 10 T (Fig.3b)
and even 20 T (not shown). Above 0.6 T, the position of the line
follows what is expected for an excitonic transition (large open
triangles in Fig.4a) in the presence of the giant Zeeman effect.
This confirms that this transition corresponds to a positively
charged exciton, with the excitation of an electron-hole pair
involving an empty $|-\frac{3}{2}\rangle$ heavy hole-subband, in
the presence of $|+\frac{3}{2}\rangle$ heavy holes.

\begin{figure}[b]
\includegraphics[width=78mm]{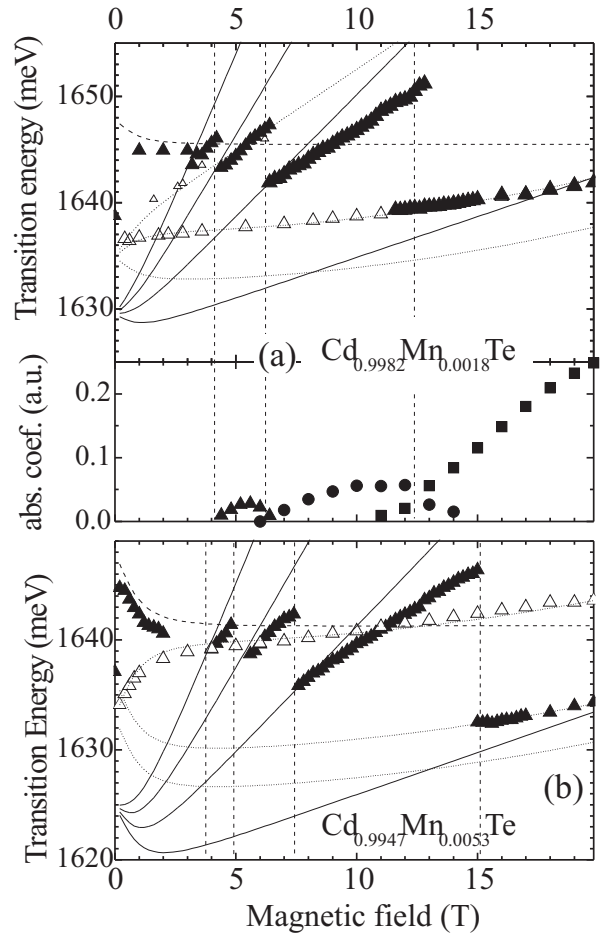}
\caption[width=85mm]{Position of the absorption lines, vs. applied
magnetic field. (a) Cd$_{0.9982}$Mn$_{0.0018}$Te QW (sample N2);
open symbols are for experimental data in $\sigma^-$ polarization,
full symbols for $\sigma^+$ polarization; dotted lines assume
excitonic-like variations (normal and giant splitting and
diamagnetic shift), with the zero-field energy as only adjustable
parameter; solid lines assume Landau levels with the theoretical
masses and the normal and giant Zeeman effects, with the zero
field energy as only adjustable parameter. The dotted line with a
higher slope is shifted from the exciton in $\sigma^-$
polarization by the Landau fan splitting (sum of cyclotron
resonances). The dashed line shows the position expected for the
absorption threshold at low field (it includes the normal and
giant Zeeman effects and the Moss Burstein shift added to the zero
field energy of the Landau fan, with no adjustable parameter). The
lower part shows the intensity of each line in $\sigma^+$
polarization. The vertical dashed lines show the magnetic field
values corresponding to integer filling factors.(b) same as (a),
for sample N5, with a Cd$_{0.9941}$Mn$_{0.0053}$Te QW.}
\label{fig04}
\end{figure}

A less intense line can be noticed at higher energy in Fig.3b, at
intermediate field (2 to 7T). It shifts linearly with the field
(small open triangles in Fig.4a). The distance between this line
and the X$^{+}$ line in $\sigma^-$ polarization is very close to
the sum of cyclotron energies of the electron and the heavy hole
(see below). Actually, this splitting is also very close to that
measured in n-doped CdTe QW \cite{Yakovlev97} between the X$^{-}$
line and a line shifting linearly with field, observed in
$\sigma^-$ polarization  while the X$^{-}$ line is seen in
$\sigma^+$ polarization, and attributed to an "exciton-cyclotron
resonance". This exciton-cyclotron resonance implies the formation
of an exciton and the promotion of an electron to the first
excited Landau level, so that the shift is determined by the
cyclotron energy of the electron (it is slightly larger due to a
recoil effect).  The same interpretation (exciton-cyclotron
resonance) cannot be applied to our case, since the splitting for
such a line in the presence of a hole gas should be close to the
cyclotron energy of the heavy hole, which is definitely smaller.
The origin of this line in our case is still unclear.

The $\sigma^+$ spectra are qualitatively different. The asymmetric
absorption line decreases as the density of $|+\frac{3}{2}\rangle$
holes increases, but the threshold can be followed up to about 3T.
Then sharper lines successively emerge, shift linearly to higher
energy (Fig.4a), and vanish (see the intensity in the lower part
of Fig.4a). Finally a single line progressively gains in
intensity. We naturally ascribe this series of lines to the
presence of the Fermi energy on successive Landau levels. The
description of Landau levels in the valence band is a complicated
matter. In the simplest approximation the in-plane effective mass
is $m_{hh}=m_{0} (\gamma_{1}-2\gamma_{2})^{-1}$. Taking Luttinger
parameters $\gamma_{i}$   from reference \cite{Dang82}, one
obtains $m_{hh}$=0.17$m_{0}$. Using a more complete model of the
valence band, a slightly larger value, $m_{hh}$=0.25$m_{0}$, was
inferred \cite{Fishman95}. This value will be used below, although
it has not been confirmed experimentally. In addition, the
transition energy is strongly affected by carrier-carrier
interactions: staying with II-VIs, examples of such effects have
been described in details for n-type CdTe and (Cd,Mn)Te QWs
\cite{Lemaitre2000}. However, the degeneracy of the Landau levels
remains unchanged and it is independent of material parameters. As
only $|+\frac{3}{2}\rangle$ holes are involved, so that each jump
occurs from one Landau level to the next one with the same spin,
we determine a filling factor $\nu$=1 at $\emph{B}$=12.5 T,
$\nu$=2 at \emph{B}=6.2 T, and so on, as shown by vertical dashed
lines in Fig.4. The carrier density in this QW is thus determined
to be \emph{p}=3.1$\times$10$^{11}$ cm$^{-2}$.

Fig.4a shows as solid lines the field dependence expected for
transitions between Landau levels, with an electron mass
$m_{e}$=0.11 $m_{0}$ and a transverse hole mass $m_{hh}$=0.25
$m_{0}$, the normal Zeeman effect, and the giant Zeeman effect
determined on the depleted QW. Using the zero field energy as the
only adjustable parameter, we obtain a relatively good agreement
for the position of the lines when they emerge at $\nu$=2, $\nu$=3
and $\nu$=4. Once again, this is a crude approximation since it
neglects the actual structure of the fan of Landau levels in the
valence band (but the larger contribution is due to the conduction
band) and carrier-carrier interactions (but these are minimal at
integer filling factors). We will show below that the zero field
energy of this Landau fan nearly matches the energy of a PL line,
which appears to be close to the band to band recombination at
\emph{k}=0. The fact that this energy of a PL line at zero field
is smaller than the excitonic transition energy is not surprising,
since the PL band-to-band transition at \emph{k}=0 leaves an
excitation in the system: the exciton initial state is lower in
energy, even if the transition energy is larger. This paradox does
not appear in transmission: Fig.4a shows (dashed line) what is
expected for the band-to-band transition at Fermi wavevector
\emph{k}$_{\mathrm{F}}$; it contains the normal and giant Zeeman
effects and the kinetic energies of the electron and the hole both
at \emph{k}$_{\mathrm{F}}$ (so that it is drawn through the middle
point between the two relevant branch of the Landau fan at each
integer filling factors). We may note that it reasonably agrees
with the position of the absorption line measured in the field
range from 1 to 3T.

Fig.4b displays the transition energies observed on sample N5,
i.e., with a slightly larger Mn content; the carrier density is
found to be \emph{p}=3.8$\times$10$^{11}$ cm$^{-2}$. Both samples
exhibit rather similar transmission spectra. We may notice that
the transmission lines in the two polarizations at filling factor
below unity are accidentally superimposed in sample N2 (a), but
are clearly separated by the larger Zeeman splitting in sample N5
(b).

\subsection{Hall effect.}
Hall effect and mobility  measurements have been performed on
several p-type doped CdTe and (Cd,Mn)Te quantum wells with a
six-contact Hall structure. Gold ohmic contacts to the buried QWs
have been obtained by growing a heavily nitrogen doped
ZnTe/(Cd,Zn,Mg)Te contact layer on top of the modulation doped
quantum well structures with a 20~nm spacer \cite{Arnoult99}.
Standard DC Hall effect and mobility measurements have been
carried out in a helium flow cryostat from 300~K down to 15 ~K and
with magnetic fields up to 1~T. The extremely high contact
resistance below 15~K, due to freezing in the barriers, did not
allow us to study the transport properties at lower temperatures.
The integrated carrier density was deduced from the slope of the
Hall resistance with respect to the magnetic field.

An example of carrier density as a function of the inverse of the
temperature was given for four different samples in Ref.
\cite{Arnoult99}. A decrease of the measured carrier density
observed between 300 and 100~K is attributed to the freezing of
the residual holes in the barrier (holes which have not been
transferred to the QW). Below 100~K and down to~15 K, a constant
carrier density is observed and attributed to the conduction of
the degenerate hole gas in the QW. In the four samples studied,
the values of the 2D hole gas carrier density vary between
2.4$\times$10$^{11}$~cm$^{-2}$ and
3.2$\times$10$^{11}$~cm$^{-2}$, in agreement with the values
deduced from optical spectroscopy (See discussion in section V).

The Hall mobility increases almost linearly up to
1000~cm$^{2}$/Vs when decreasing the temperature from 300~K down
to 15~K. This leads to a typical mean free path of the 2D hole gas
of the order of 10~nm. This is a lower bound however since we
observed a strong broadening of the PL spectra of the samples
prepared for Hall studies, which had to be glued on the sample
holder and appear to be highly strained.

\section{Photoluminescence: band-to-band vs. charged exciton.}
This section is devoted to the description of the PL spectra and
the effect of the Zeeman splitting: first, in A, we show the
destabilization of the charged exciton - in favor of the neutral
exciton - in a weakly doped QW. Then, in B, we show that a similar
effect occurs in a QW with a larger carrier density. Finally, we
describe more precisely the case of a partially polarized carrier
gas in C.

\subsection{Low concentration: charged and neutral excitons.}

\begin{figure}[b]
\includegraphics[width=85mm]{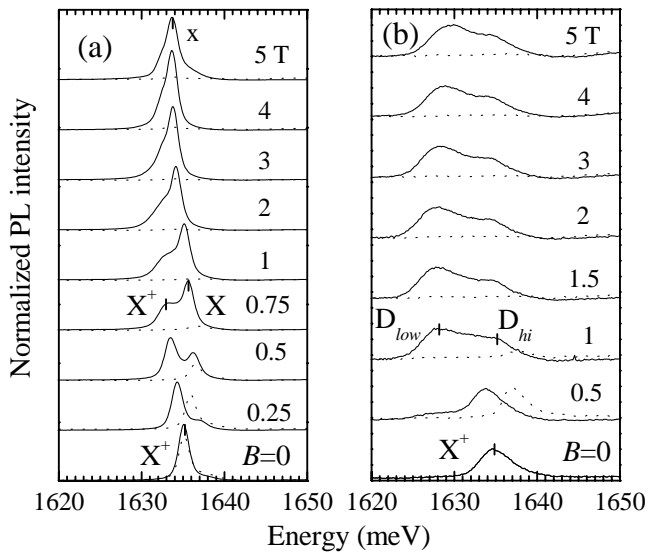}
\caption[width=85mm]{(a) PL spectra, at 1.7 K, with various values
of the applied field, from a Cd$_{0.9963}$Mn$_{0.0037}$Te QW
(sample N3); the PL was excited using an Ar-ion laser which almost
completely depletes the QW . Solid lines are for $\sigma^+$ spectra
and dotted line for $\sigma^-$; (b) same as (a), but with
excitation with a Ti-sapphire laser with photon energy below the
barrier gap, which has no effect on the density of the carrier
gas.}
\label{fig05}
\end{figure}

\begin{figure*}[t]
\includegraphics*[width=160mm]{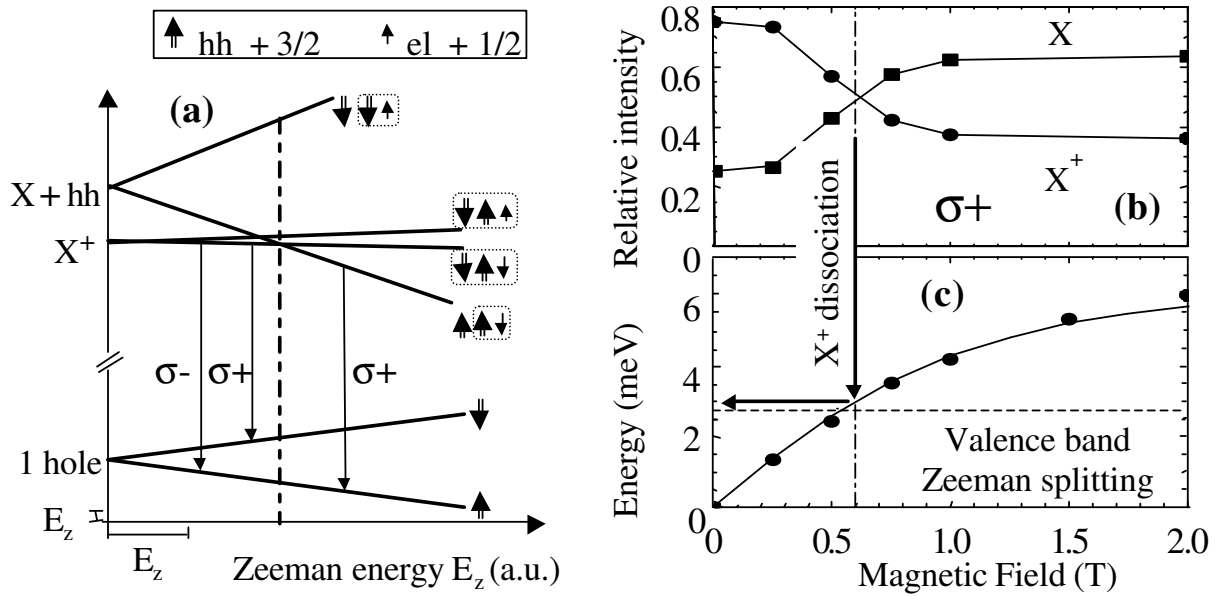}
\caption[width=140mm]{(a) Three-carriers diagram of optical
transitions, single arrows show the electron spin and double
arrows the hole spin; the dash-dotted line marks the level
crossing. (b) relative intensities of the X and X$^{+}$
photoluminescence lines measured in $\sigma^+$ polarization for
sample N3, Cd$_{0.9963}$Mn$_{0.0037}$Te QW, under illumination by
Ar ion laser, (c) comparison of the valence band Zeeman splitting
(points and solid line) and the X~/~X$^{+}$ splitting (dashed
horizontal line) for sample N3. Dot-dashed vertical lines mark the
field at which the X~/~X$^{+}$ crossing occurs. } \label{fig06}
\end{figure*}

We start the  discussion of the PL spectra by the limit of low
carrier density. Fig.5a shows PL spectra at different values of
the magnetic field, in both circular polarizations, observed on
sample N3 with 0.37$\%$ Mn in the QW (which allows us to follow
the PL lines over the whole field range without overlap with the
PL from the substrate). The carrier density was decreased by
illumination with an Ar laser (which at the same time excites the
PL) and can be estimated to be below 2$\times$10$^{10}$~cm$^{-2}$
in the conditions of Fig.5a. The identification of the lines is
based on the comparison between PL spectra and transmission
spectra with white light (not shown). At zero field, the spectrum
is dominated by a single line related to the charged exciton. A
much weaker line due to the neutral exciton appears as a shoulder
on the high energy side. When applying the magnetic field, the
relative intensity of both lines changes. In $\sigma^-$
polarization, the neutral exciton line (the shoulder) disappears
completely above 0.3~T, and the intensity of the charged exciton
line progressively decreases; it disappears at 1~T. Fig.~6b shows
the intensity of each line. In $\sigma^+$ polarization, the PL
intensity of the neutral exciton increases with respect to the
X$^{+}$ one. The intensities of both lines become comparable at
0.6~T, and at higher fields the neutral exciton line dominates the
spectra.

\begin{figure}[ht]
\includegraphics[width=83mm]{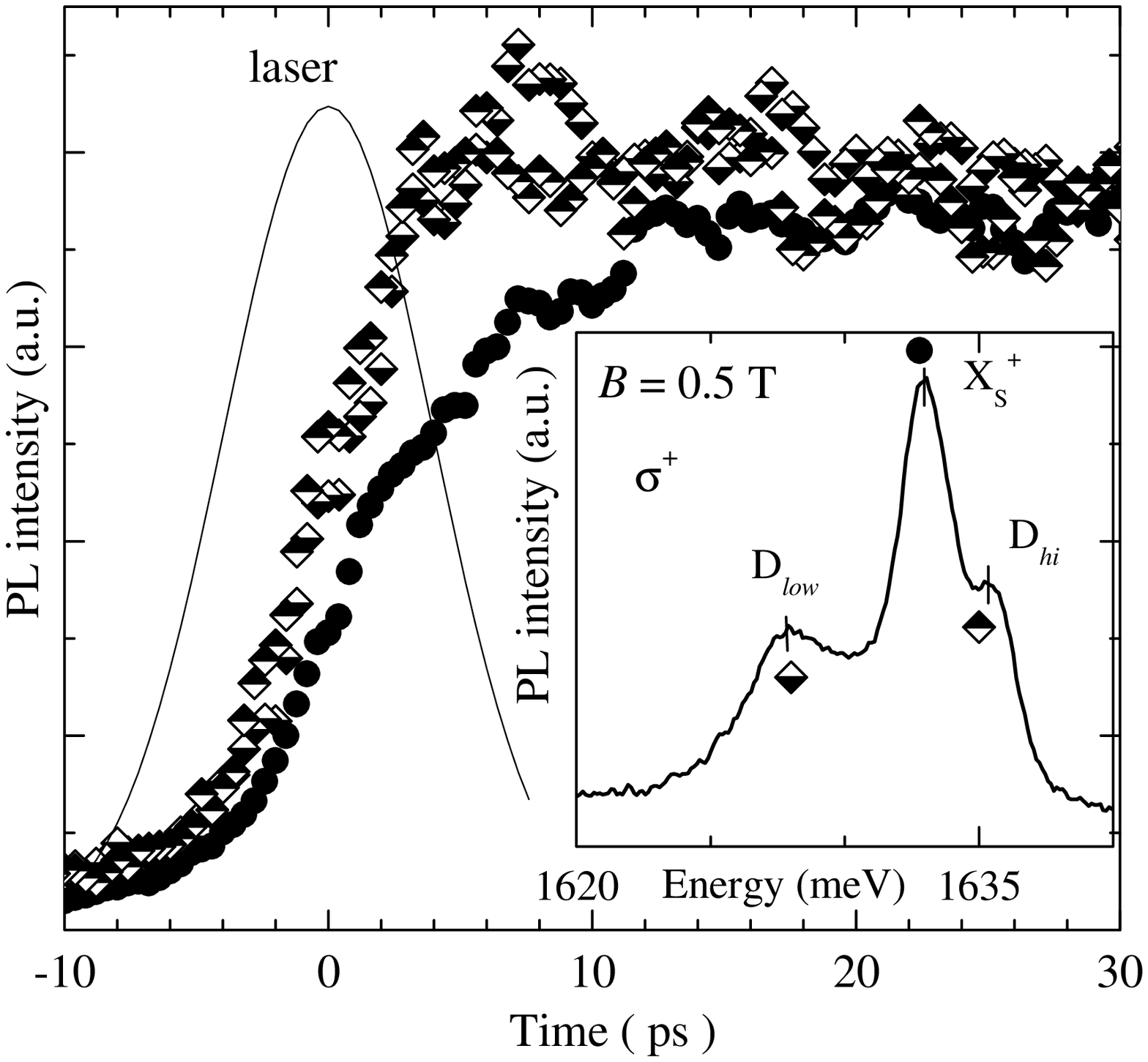}
\caption[]{ Time-resolved photoluminescence of sample N4
(Cd$_{0.996}$Mn$_{0.004}$Te QW) measured in $\sigma^+$
polarization in a magnetic field of 0.5~T close to the crossing of
the excited levels. Symbols give the temporal profiles after
excitation by the laser pulse (solid line), detected at different
energies corresponding to different components of the PL (marked
D$_{low}$, X$_{s}$$^{+}$, D$_{hi}$ on the spectrum in the inset).}
\label{fig07}
\end{figure}

This destabilization of the charged exciton in favor of the
neutral exciton can be understood as the result of the different
Zeeman splittings of the three-particle system (two holes and an
electron) either arranged as a charged exciton, or containing a
neutral exciton. A schematic diagram of the three-particle levels
is shown in Fig.6a. First, we have checked on samples with
different Mn contents (not shown) that the transition from the
charged to the neutral exciton is governed by the value of the
giant Zeeman splitting. When applying a magnetic field the two
states of the three-particle system experience different giant
Zeeman shifts. In the fundamental state of the charged exciton,
the two holes are arranged in a singlet state, therefore the
Zeeman shift of the three-particle system is equal to the shift of
the electron only, which is rather small: in (Cd,Mn)Te, the giant
Zeeman shift of the heavy hole is four times larger that the
electron shift. The bright state of the neutral exciton is
composed of a hole and an electron with opposite spin directions,
and the remaining hole is free; therefore the component with its
two holes in the majority band (which emits in $\sigma^+$
polarization) shows a strong redshift under applied field, equal
to the sum of the individual shifts of the electron and the two
holes. At zero field, the charged exciton state is lower in energy
since it is a bound state, and this remains true at low field.
When the heavy-hole Zeeman splitting is equal to the X$^{+}$
dissociation energy (Fig.6a,c), the neutral exciton level crosses
the level of the charged exciton emitting in the $\sigma^+$
polarization, so that the neutral exciton level becomes lower in
energy. The mechanism is similar to the crossing induced by the
giant Zeeman splitting between the A$^{0}$X bound excitons and
free excitons in bulk diluted magnetic semiconductors
\cite{Planel80}.

The ratio of the X and X$^{+}$ intensities experimentally observed
(Fig.6c) varies smoothly, making it difficult to precisely point
out the crossing. Actually, the intensity ratio results from the
dynamics between the X and X$^{+}$ states, and, in addition,
disorder introduces a broadening of the densities of states. It
was already recognized \cite{Vanelle00} that the dynamical
non-equilibrium distribution between the X and X$^{+}$ populations
leads to the observation of some PL intensity from the upper
level. In particular, at zero field, the neutral exciton
luminescence is observed due to a formation time of the charged
exciton comparable with the decay time of the neutral exciton.
This formation time is strongly sensitive to the carrier density
\cite{ Kossacki2003}: As the carrier density is increased, it
becomes shorter so that a more abrupt transition is expected.

The main result of this paragraph is that the giant Zeeman
splitting of the hole can be large enough to destabilize the
singlet state of the charged exciton, in favor of the $\sigma^+$
emitting neutral exciton, which is such that the two holes have
parallel spins.

\subsection{Photoluminescence in the presence of a 2D hole gas.}
The main features of the PL spectra at moderate Zeeman splitting
have been described in \cite{Kossacki99}. It was shown that, in
$\sigma^-$ polarization, the PL line and the absorption line
coincide (but for a constant Stokes shift smaller than 1~meV) over
a well defined range of the applied field where the hole gas is
fully polarized; it was shown also that in the same field range
the PL line in $\sigma^+$ polarization is shifted by the excitonic
Zeeman splitting. This suggests that both lines are due to the
charged exciton. We will come back to this point - and to the
shift which appears between the absorption and PL lines at smaller
field - in section IV-C. First we keep in mind this assignation of
the PL lines at low Zeeman splitting to the charged exciton and we
wish to introduce a new feature which appears at larger Zeeman
splitting.

\subsubsection{Appearance of a double line at large Zeeman splitting.}
Fig.5b shows PL spectra, at different values of the applied
magnetic field in both circular polarizations, observed on the
same sample N3 as discussed before, but measured with Ti-sapphire
laser excitation, hence without above-barrier illumination. The
carrier density will be estimated as explained below as
\emph{p}=4.2$\times$10$^{11}$~cm$^{-2}$. Similarly to the lower
density case, a single line is observed at zero field. At low
field (see the spectra at 0.5~T), the giant Zeeman effect is
sizable; the high energy line is polarized $\sigma^-$ and sharp,
while the low energy line, polarized $\sigma^+$, is broader, with
a weak tail on the low energy side. We continue to assign this PL
to the charged exciton \cite{Kossacki99}, and we will come back to
this later. At higher field, the PL is fully polarized $\sigma^+$,
with a typical double structure containing a high energy line
labelled D$_{hi}$ in the following, and a low energy line labelled
D$_{low}$. The transition from the single to the double line in
$\sigma^+$ polarization occurs at about 0.6~T, close to the value
causing the X/X$^{+}$ crossing at low carrier density. The shape
of the double line remains constant up to 3~T in Fig.5b. Above
this field a progressive change of shape occurs, which we
attribute to the emergence of the Landau levels as seen in Fig.4.

We will discuss this evolution in the same spirit as the one
observed from charged to neutral excitons for low carrier density.
More precisely, we will show that the double line involves a
unique initial state, which crosses the charged exciton state when
the Zeeman splitting exceeds some value.

\begin{figure}[h]
\includegraphics[width=83mm]{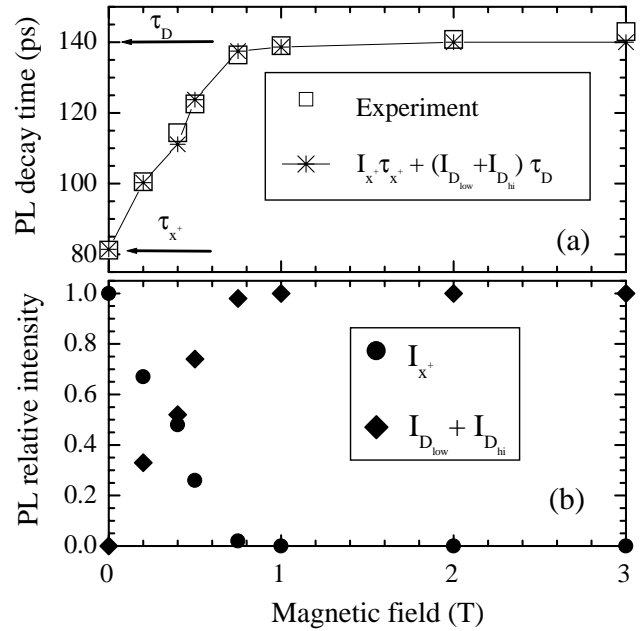}
\caption[]{Sample N4, Cd$_{0.996}$Mn$_{0.004}$Te QW (a) integrated
PL decay time measured in $\sigma^+$ polarization, as a function
of the magnetic field; open squares denote the experimental
values, stars represent the average of two times:
$\tau_{X^{+}}$=80~ps and $\tau_{D_{low}}$=$\tau_{D_{hi}}$~140ps
(arrows), weighted by the intensities of each of the two
components of the PL spectra, plotted in (b); (b) relative
time-integrated intensities of the two components of
photoluminescence spectra: double line D$_{low}$ / D$_{hi}$
(squares), and excitonic line (circles) measured in $\sigma^+$
polarization, as a function of the magnetic field.} \label{fig08}
\end{figure}

First, the character of the optical transition involved in this
double line was examined by time resolved spectroscopy. The inset
in Fig.~7 shows a time-integrated spectrum taken in $\sigma^+$
polarization in a field of 0.5~T. This is the field range for
which both the single and the double lines are observed, and the
hole gas is already fully polarized. PL was excited with a ps
pulse also with $\sigma^+$ polarization. The decay time of all
three lines was found to be the same, about 100~ps. Rise times
however were quite different. PL traces related to the double line
(traces D$_{hi}$ and D$_{low}$ in Fig.7) exhibit the same rise
time, faster than 2~ps. A significantly larger rise time, larger
than 3~ps, is measured on the single line (trace X$_{s}$$^{+}$ in
Fig.7). When increasing the magnetic field the rise times of both
components of the double line remain the same and very fast. This
we use as a first argument that the two components D$_{hi}$ and
D$_{low}$ of the double line involve the same initial state, which
is different from the initial state of the single line observed at
lower field.

The values of the rise times also support the assignation of the
single line to a state involving a two-hole singlet, and the
double line involving an initial state where all holes have the
same orientation as in the ground state. The $\sigma^+$ polarized
light creates an electron-hole pair where the hole has the same
spin orientation as those of the carrier gas (majority spin).
Hence the charged exciton luminescence involves a heavy hole spin
flip in order to form the singlet hole pair. In fact the time of
3~ps compares well to the heavy hole spin flip time observed in
X$^{-}$ states \cite{Kossacki2000,Ciulin2002}. On the other hand,
the fast rise of the double line suggests that no spin flip - and
hence no singlet hole pair - is involved in this double line.

The picture of a crossing of initial states is further supported
by the variation of the decay time versus magnetic field. Fig.8a
presents the dependence of the integrated PL decay time versus
magnetic field. This time increases by almost a factor of 2 when
increasing the magnetic field from 0 to 1~T, which coincides with
the transition from the single PL line to the double line. The
observed decay time coincides exactly with the average of the
decay times at 0~T and 3~T, weighted by the relative intensities
of the two PL components, presented in Fig.8b. Such a behavior is
predicted by a rate equation model if there are only two
field-independent channels for radiative decay, related to the two
PL mechanisms giving rise to the single and double lines,
respectively. The double line is associated to a decay channel
(140~ps) slower than that of the charged exciton line (80~ps),
suggesting a weaker correlation.

We conclude that the transition between the two kinds of PL
spectra (excitonic or double-line) is related to the level
crossing of the initial states, with one state involving a pair of
holes forming a singlet (charged exciton recombination at low
field), the other one with all hole spins having the same
orientation. Here again, the field at which the transition takes
place is defined by a value of the valence band Zeeman splitting
(see below the compilation over various samples), and the
transition will be more or less abrupt, depending on the
relaxation processes between the two levels. It is worth
mentioning that, as expected, the transition is more abrupt in cw
experiment with very weak excitation, than in time resolved
experiment for which the system is excited farther from
quasi-equilibrium conditions.

\begin{figure}
\includegraphics[width=83mm]{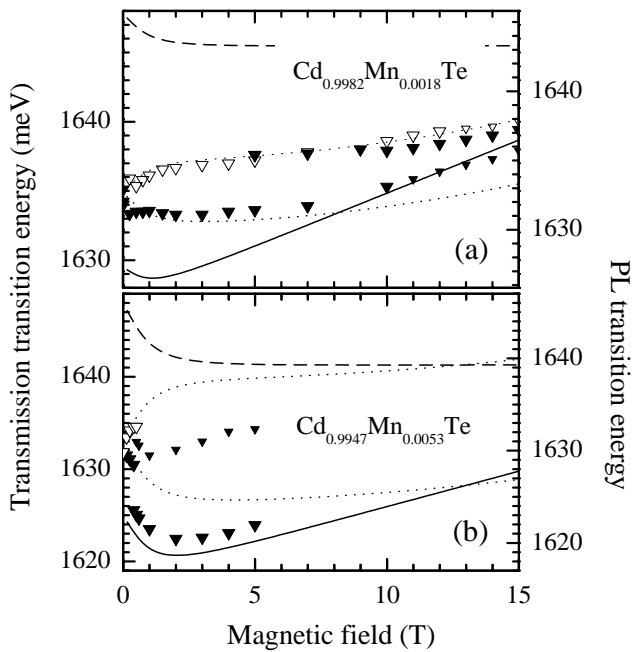}
\caption[]{Comparison of the transition energy in PL (down
triangles) versus transmission (lines showing the fits used in
Fig.4). Note that the PL and transmission scales have been shifted
(see text). (a) sample N2, Cd$_{0.9982}$Mn$_{0.0018}$Te QW and (b)
sample N5, Cd$_{0.9947}$Mn$_{0.0053}$Te QW, as in Fig.4}
\label{fig09}
\end{figure}

\begin{figure*}
\includegraphics*[width=175mm]{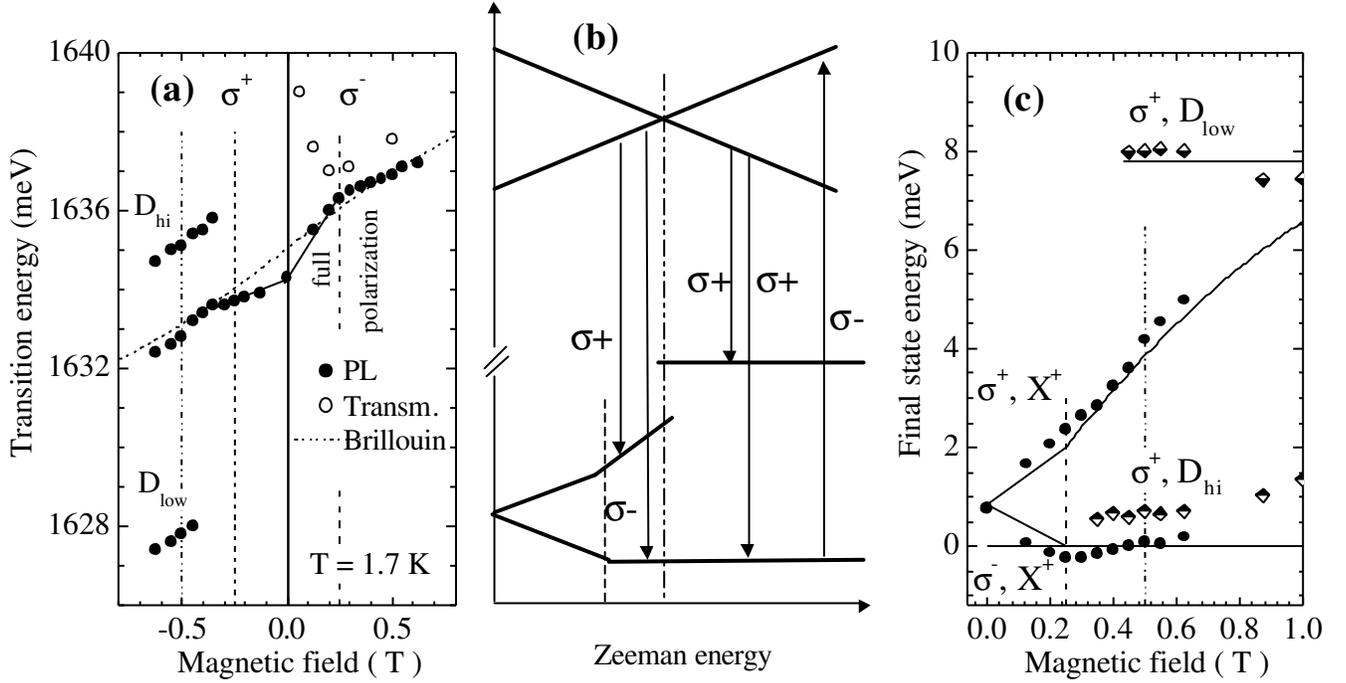}
\caption[]{(a) Positions of the transmission lines (open circles)
and PL lines (full circles), vs. the applied magnetic field, in
the low field range, for sample N4, Cd$_{0.996}$Mn$_{0.004}$Te QW.
The $\sigma^-$ data are plotted at negative fields. The dotted
line gives the Brillouin function determined at high field (b)
schematic diagram of the initial and final states and of the
optical transitions, (c) experimentally determined energies (see
text) of the final states of optical transitions, obtained from
data in (a); diamonds are for the double line
(D$_{low}$,D$_{hi}$), circles for the excitonic line and the solid
lines are drawn using the energies of initial and final state
schematized in (b). In all 3 panels, dashed vertical lines
indicate the field values for complete hole polarization, and the
dash-dotted vertical line indicates the destabilization of X$^{+}$
.} \label{fig10}
\end{figure*}

Now we come back to the analysis of the energies of the PL lines
in sample N5, for field values such that the hole gas is fully
polarized (above 0.2T in this sample, see below). Fig.9b compares
the positions of the transmission lines (the fits from Fig.4b), to
the positions of the PL lines in the same sample. Transmission
spectra were measured in one run at the Grenoble High Field
Laboratory while we used a Ti-sapphire laser to excite the PL in a
different run: to take into account the expected Stokes shift
(usually at most 1~meV when transmission and PL were measured in
the same run) and also to allow for a possible fluctuation of
sample characteristics between the regions of the sample observed
during the two runs, the PL scale was shifted (by 2 meV), so that
PL and transmission lines in $\sigma^-$ polarization were made to
coincide at one field where both are observed. As shown previously
\cite {Kossacki99} for sample N2, and below in Fig. 10a for sample
N4, this plot confirms that the PL transition in $\sigma^-$
polarization closely matches the position of the charged exciton
measured in transmission over a finite field range; but that
appears to be true also for the $\sigma^+$ PL at low field (below
0.3~T in this sample). Above this field value, the low energy
component of the PL double-line (D$_{low}$) is close to the
position expected for the band-to-band transition (lower branch of
the Landau fan). The high energy component of the $\sigma^+$
doublet (D$_{hi}$) is marked with smaller symbols because its
intensity is small in this sample (in addition at this particular
Mn content it overlaps with the substrate PL so that its energy
was determined with less accuracy). Note also that the difference
in radiative decay times observed on sample N4 of Fig.8a (80~ps
and 140~ps respectively) is consistent with such an excitonic
character of the single dichroic line at low field vs. a
band-to-band character of the double line D$_{low}$-D$_{hi}$: The
excitonic decay time is expected to be significantly shorter than
the one measured in the band-to-band transition which implies a
weaker electron-hole correlation.

The jump to the double-line is induced by the giant Zeeman
splitting. This was checked on all samples with a Mn content
between 0.2 and 1$\%$. For example, in sample N2 of Fig. 4a, with
the lowest Mn content of our series, the jump is not observed
(Fig.9a) up to 10~T, although the hole gas is fully polarized
already at 0.6~T. A progressive change in the slope of the
$\sigma^+$ PL energy versus field would agree with the same change
of nature of the PL transition, from excitonic to band-to-band,
but Landau levels cannot be ignored at such high fields. In this
sample with a very low Mn content the maximum Zeeman splitting is
large enough to fully polarize the hole gas, but too small to
destabilize the charged exciton singlet state.

Let us briefly summarize our findings at this point. In the
presence of a fully polarized hole gas with a density of several
10$^{11}$~cm$^{-2}$, PL spectra exhibit a crossing of the excited
states which behaves quite similarly to the X/X$^{+}$ crossing
observed at low hole density. At fields below the crossing, we
observe a single PL line in both $\sigma^+$ and $\sigma^-$
polarizations. Its position coincides with the charged exciton
X$^{+}$ (which is observed to emerge in $\sigma^-$ transmission
from the charged exciton when the carrier density increases) and
it has a short decay time (as expected for an exciton) and a long
formation time (as expected for the singlet configuration of the
two holes in the charged exciton). Thus, the initial state of this
transition is likely to be linked to the charged exciton state
well identified at low carrier density. Above the crossing, the
giant Zeeman energy due to the two-hole singlet is too large. The
excited state which is now lower in energy gives rise to a double
PL line, D$_{low}$-D$_{hi}$; it does not contain a two-hole
singlet, so that its formation time is shorter, and the
electron-hole correlation is weaker so that the lifetime is
longer. Finally, the position of the low-energy component
D$_{low}$ is close to what we would expect for a band-to-band
transition at \emph{k}=0.

The crossing of singlet and triplet levels plays also an important
role in the polarization of the total PL signal. The two-hole
singlet states which are the initial states of PL transitions in
both circular polarizations differ only by the spin of the
electron. Thus the splitting of this level is only 1/5 of the
exciton Zeeman splitting. Such a small splitting leads to
comparable occupations of both states and hence to comparable PL
intensities in both polarizations. Additionally, the spin flip
time of the electron is relatively long and even comparable to the
X$^{+}$ lifetime \cite{Vanelle00,Kossacki2003}. This prevents fast
spin relaxation between the charged exciton singlet states. A
completely different situation takes place after singlet-triplet
crossing. Then only the $\sigma^+$ transition is possible from the
lowest initial state. The opposite triplet state has an energy
higher by twice the value of the exciton splitting. This is in
agreement with the PL polarization experimentally observed: The
$\sigma^-$ line intensity remains almost the same as that in
$\sigma^+$ below the jump, and it vanishes completely after the
jump.

The conclusion of this paragraph is that the giant Zeeman effect
induces a crossing between an excited state which involves a
two-hole singlet and another one where all holes are in the
majority spin subband. There are strong hints that the first state
is the singlet state of the charged exciton X$^{+}$; the second
one could be an uncorrelated electron-hole pair (initial state of
band-to-band transitions) or a weakly correlated "triplet state".
With these attributions in mind, we will continue the analysis of
the spectra.

\begin{figure}[]
\includegraphics[width=85mm]{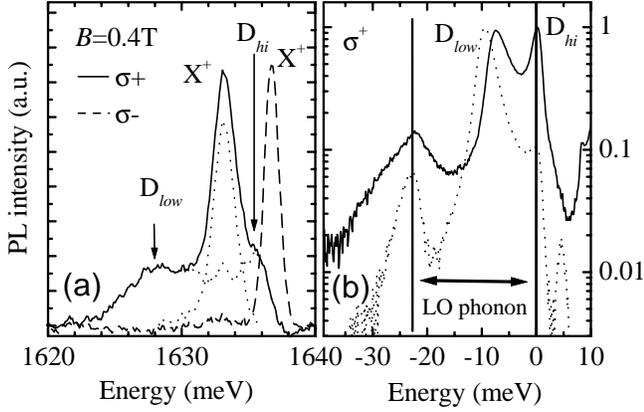}
\caption[]{(a) cw PL spectra of sample N4 at level crossing
(linear scale); the solid line gives the experimental PL spectrum
in $\sigma^+$ polarization and the dashed line the experimental
spectrum in $\sigma^-$ polarization; the dotted lines give the
proposed decomposition of the $\sigma^+$ line into the double line
(D$_{hi}$-D$_{low}$) and the excitonic one; (b) typical spectra in
$\sigma^+$ polarization for samples N4 (at 1~T) - solid line, and
S7 (QW doped from surface states, at 0.5 T) - dotted line; the
energy scale is with respect to the position of the upper
component of the PL double line, D$_{hi}$, and the intensity scale
is a log scale.} \label{fig11}
\end{figure}

\subsubsection{Final states in PL transitions}

We now discuss the final states assuming the same field dependence
of the initial states as in the low density case (dependence
determined by the spin singlet or spin triplet configuration of
the two holes). We will show that each of these initial states
gives rise to two transitions: one toward the ground state, and
another one leaving the hole gas in an excited state. In order to
describe the PL in the case of a large density of the hole gas, we
redraw the three-particle scheme introduced to describe X and
X$^{+}$ (Fig.6a), now measuring all energies with respect to the
ground state of the hole gas (Fig.10b).

The simplest transition takes place in $\sigma^-$ polarization at
sufficiently high magnetic field, so that the hole gas is fully
polarized. For each field, below the destabilization of the
singlet state where $\sigma^-$ PL disappears, both PL and
absorption exhibit a single $\sigma^-$ sharp line, at almost the
same energy (but for a small, constant, Stokes shift). Therefore
we attribute this line to the transitions between the ground state
of the carrier gas and some exciton complex. Due to the selection
rules in transmission this complex is identified as a charged
exciton with a two-hole spin singlet and a +$\frac{1}{2}$
electron. In ($\sigma^-$) PL, the +$\frac{1}{2}$ electron
recombines with the minority (upper energy) -$\frac{3}{2}$ hole of
the singlet, leaving all the holes in the lower, majority spin
subband. In $\sigma^+$ polarization, the absorption transition
involving the singlet charged exciton state is forbidden. However
if the electron spin flips to -$\frac{1}{2}$, the PL transition is
allowed in $\sigma^+$ polarization, but does not lead to the
ground state of the hole gas: The -$\frac{1}{2}$ electron
recombines with a +$\frac{3}{2}$ hole, leaving one -$\frac{3}{2}$
hole in the opposite subband. Therefore the final state is an
excited state of energy equal at least to the hole Zeeman
splitting.

When the two excited states cross each other, the initial state of
the transition changes its character. Experimentally, a PL jump is
observed in $\sigma^+$ polarization, where the single PL line
turns to a double one. We now identify the final states of the
transitions related to the double line by considering the
crossing. At this particular field, the initial states in
$\sigma^+$ polarization are degenerate, and the only difference
between the two singlet initial states in $\sigma^+$ and
$\sigma^-$ polarization is the Zeeman splitting of the electron.
We already know that the final state in $\sigma^-$ is the ground
state of the hole gas. According to our measurements for all
samples, at the crossing, the upper component of the PL double
line, D$_{hi}$, has almost the same energy as the $\sigma^-$
transition (see Fig.11a): the difference, around 1~meV, is partly
due to the electron Zeeman energy (of the order of 0.5~meV at the
jump), and what remains might be due to the different shapes of
the two lines. Therefore the final state of D$_{hi}$ is also the
ground state of the heavy hole gas, while D$_{low}$ leaves it in
an excited state.

Note that in a band-to-band description of this double line, the
transition D$_{low}$ is a direct one involving an electron and a
hole both at \emph{k}=0 (thus leaving the hole gas in an excited
state of energy equal to the Fermi energy of the spin-polarized
gas, 2\emph{E}$_{\mathrm{F}}$, keeping the notation
\emph{E}$_{\mathrm{F}}$ for the Fermi energy at zero field), while
D$_{hi}$ is an indirect one where the electron recombines with a
hole at Fermi level, leaving the hole gas in the ground state.
This indirect transition might be allowed due to disorder or many
body effects. In fact we observe that the relative intensity of
this line D$_{hi}$ with respect to the lower one D$_{low}$ varies
from sample to sample. The indirect character is further supported
by the observation of the LO-phonon replica: In magnetic field in
$\sigma^+$ polarization, a clear phonon replica is observed only
for D$_{hi}$ (Fig.11b). This is characteristic for transitions
which are indirect in \emph{k}-space, so that the replica includes
the phonon wave vector in the conservation rule. From a purely
experimental point of view, the splitting of the double line gives
the energy of the excited state of the hole gas towards which the
transition takes place in the case of the lower component. The
variation of this energy versus carrier concentration will be
discussed later in section V.B.

The ordering of the transition energies may be different from the
ordering of the initial states involved in the transition: the
difference is due to the energy of the final state, which of
course is not necessarily the ground state. Hence, an excitonic
transition may appear at an energy higher than a band-to-band
transition. This is the case here if we compare the
charged-exciton line in $\sigma^-$ polarization (which has the
ground state as the final state) and the intense low-energy
component of the double-line, which leaves the hole gas in an
excited state.

The hole Zeeman splitting, for destabilization field (the field at
which the level crossing takes place), is a measure of the binding
energy of the singlet state at zero field. We found this energy to
lay between 2~meV and 3~meV for all measured samples,
independently of the carrier density (figure 13).

Summarizing this part, the PL has an excitonic character for
$\sigma^-$ polarization, and in $\sigma^+$ at low field. At high
field in $\sigma^+$ polarization, a double line is observed. This
change is due to a level crossing between the charged exciton
state which contains two holes in a singlet configuration, and an
initial state of the double line where all holes are in the
majority spin subband: due to the strong Zeeman energy, the hole
pair involved in the charged exciton flips from the singlet
configuration to a triplet. Several characteristics of this double
line (including its shape, but also the dynamics and the
coincidence in energy of the lower component with the energy
between first Landau levels extrapolated to \emph{B}=0) suggest
that this transition is quite similar to a band-to-band
transition. The upper component of the double line involves the
transition to the fundamental state of the hole gas. In a
band-to-band picture, such a transition is an indirect one since
the electron recombines with a hole at Fermi level. The lower
component of the double line is related to a transition towards an
excited state of the hole gas. This excitation involves no spin
flip (all holes are in the majority spin subband). In a
band-to-band picture, it is the state after recombination at
\emph{k}=0, in which one electron is left at the top of valence
band.

\subsection{Low fields - incomplete hole gas polarization}
For low magnetic fields, the dependence of the PL energy on
magnetic field significantly deviates from the curve describing
the giant Zeeman effect in (Cd,Mn)Te. An example is given in
Fig.~10a for sample N4. The transition energy measured at zero
field is 1~meV below the Zeeman curve (dotted line in Fig.~10a).
This difference decreases for both polarizations when increasing
the magnetic field, and vanishes for a field stronger than a
certain value, equal to 0.2~T in sample N4. Above this field, the
PL line in $\sigma^-$ polarization coincides with the absorption
line (but for a small constant Stokes shift). This lets us
conclude that this point is a point of complete polarization of
the hole gas.

This can be used to obtain an information on the final state
involved in the transition. This is done by assuming, as above,
that the energy of the initial state (the charged exciton)
involves only the electron Zeeman shift, so that the energy of the
final state (fig.~10c) is deduced from the transition energy. As
expected, in $\sigma^-$ polarization, the final state is the
ground state of hole gas. In $\sigma^+$ polarization, the same
transition leads to an excited state of the hole gas, with one
hole transferred from the majority spin subband of the
valence-band, to the empty spin subband. This excited state is a
spin excitation of the hole gas. The k selection rule is obeyed if
this excitation is a spin wave at k=0 or, in an
independent-carrier description, there is one electron at k=0 in
the majority spin subband and a hole at k=0 in the empty spin
subband: the excitation energy is equal to the Zeeman splitting of
the valence band, 2\emph{E}$_{\mathrm{Z}}$ where
\emph{E}$_{\mathrm{Z}}$ is the Zeeman shift, as experimentally
found (fig.10c, above 0.2 T).

Similarly, we can calculate the energies of the final state also
for the incompletely polarized hole gas. We still assume that the
electron is left at the top of the valence band. Then the energy
of the final states are
\emph{E}$_{\mathrm{F}}$+\emph{E}$_{\mathrm{Z}}$ for $\sigma^+$
polarization and \emph{E}$_{\mathrm{F}}$-\emph{E}$_{\mathrm{Z}}$
for $\sigma^-$ polarization, where $E_\mathrm{F}$ is the Fermi
energy of the hole gas measured from the top of  the valence band
at zero field. These energies are plotted schematically in
Fig.~10b, and compared to the experimental values below 0.2 T in
Fig.~10c. Fig.~10c shows also that, as described in section
IV-B-2, the final states of the double line are, to a good
approximation, the ground state and an excited state of
field-independent energy.

It is interesting to note that according to the proposed level
diagram, the distance from the charged exciton line to the upper
component of the double line in $\sigma^+$ polarization is equal
to the binding energy of the singlet. Thus, this binding energy
(which was deduced above as being equal to the Zeeman splitting at
the field where the jump occurs) can be measured directly on the
spectra at the jump, as the splitting between two lines.

\begin{figure}[h]
\includegraphics[width=85mm]{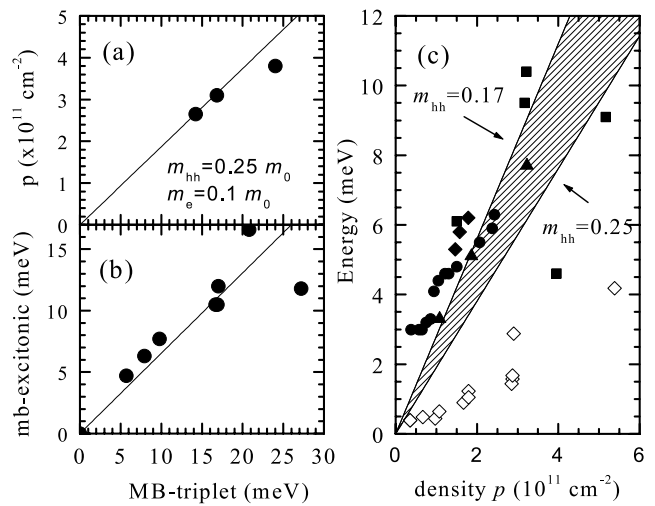}
\caption[]{(a) The hole gas density determined from filling
factors in the Landau fan, or from Hall resistance, versus the
polarized Moss-Burstein shift between the absorption line and the
lower component of the double PL line, D$_{low}$, in $\sigma^+$
polarization; the solid line is the calculated slope using the
electron and hole masses as described in the text, (b) "excitonic
mb-shift" (measured in $\sigma^+$ polarization between the
absorption line and the excitonic PL line, at a field such that
the hole gas is completely polarized, but which is lower than the
value where the level crossing occurs. The solid line is a linear
fit (c) Full symbols represent the splitting of the double line
D$_{hi}$ - D$_{low}$ - the energy of the final state involved in
the D$_{hi}$ component -versus the carrier density. Note that two
samples have been measured at different carrier densities,
changed either optically (sample N4, triangles) or through en
electrostatic gate (sample D3, circles and sample D50, diamonds).
Open diamonds represent the valence band Zeeman splitting at
complete hole gas polarization; grey area: Fermi energy of the
polarized hole gas.} \label{fig12}
\end{figure}

\section{Characteristic energy scales of the system.}

This section is devoted to discussion of characteristic energies
related to the carrier gas. First, in A, we compare different
methods of determination of the carrier density. Then in B, we
discuss the energies of excitations of the carrier gas, the Zeeman
energy necessary to completely polarize the hole gas, and its
Fermi energy. Finally, in C, we describe the the evolution of the
charged exciton dissociation energy with carrier density.

\subsection{Spectroscopic determination of the carrier density}
In this paragraph we wish to complete our set of methods of
characterizing the hole gas in a QW. The carrier density is often
determined from the shift between the PL and PLE or absorption
lines: this is the so-called Moss-Burstein shift, which in the
case of band-to-band transitions is the sum of the kinetic
energies of holes and electrons at \emph{k}$_{\mathrm{F}}$. To
increase the accuracy, these measurements can be carried out in a
magnetic field, in which the hole gas is completely spin
polarized, so that the magnitude of the shift is doubled. Here we
profit from the samples for which the carrier concentration was
determined by direct techniques, Hall effect in one sample,
filling factors at high field in two samples. These three samples
were also characterized by standard magnetooptics, and the
so-called Moss-Burstein shift was obtained.

First we compare the values of the density obtained from filling
factors of Landau levels in transmission and from Hall
measurements, to the shift between the lower component of the
double line in PL, D$_{low}$, and the absorption in $\sigma^+$
polarization (Fig.13a). The slope is very close to the value
expected for the Moss-Burstein shift $E_{\mathrm{MB}}$ in a single
particle approach,
$$
p=\frac{m_{0}}{2\pi\hbar^{2}}\frac{m_{e}m_{hh}}{m_{e} + m_{hh}}E_{\mathrm{MB}}.
$$
This confirms that the double line is very close to a band-to-band
transition. Therefore this is in principle the better choice for a
spectroscopic determination of the carrier density.

However, the shift between the charged exciton PL and the
absorption line, at intermediate field, is usually much easier to
measure. Here again, it can be measured in a magnetic field high
enough to fully polarize the hole gas, but below the jump to the
double line \cite{Kossacki99}. We will call this shift "the
excitonic mb shift". We find it to be a linear function of the
real Moss-Burstein shift and of the carrier density (Fig.12b). In
the present case this "excitonic Moss-Burstein shift" is smaller
than the calculated Moss-Burstein shift by a factor of 1.5. Hence
we confirm that this procedure \cite{Kossacki99} provides an
accurate evaluation of the relative hole density, and even of its
absolute value once a proper calibration has been done.

The final state in the $\sigma^+$ absorption process has to be
discussed, probably by considering an excitation of the hole gas
in the presence of the bound state (singlet trion). The intensity
of this line rapidly decreases in $\sigma^+$ polarization, since
it needs the presence of a minority spin hole to form the two-hole
singlet state. Note that, for this reason, a clear distinction has
to be made between the excitonic transition, observed in the
presence of the incompletely polarized hole gas, and the band to
band transition, which does not include a singlet state and is
observed (see Fig.~4) at field values up to a few teslas. However,
the intensity of the two absorption lines, the charged exciton
transition in $\sigma^+$ polarization and the band-to-band
transition, is too low, in a single QW sample, to allow a detailed
study.

To sum up, we have several well-justified determinations of the
carrier density (filling factors, Hall effect, MB shift between
D$_{low}$ and transmission), but the most convenient one is to
measure the "excitonic mb shift" between the charged exciton
luminescence and the transmission line. This can be done at zero
field, or, with a better accuracy, at moderate field; a
preliminary calibration is needed (Fig.12). It implies that the
carrier densities quoted in our previous studies \cite{Kossacki99}
were underestimated by a factor of 1.5.

One could think that the Zeeman effect needed to fully polarize
the hole gas could be another way of determining its density. We
will show below that this involves a strong contribution from
carrier-carrier interactions

\subsection{Energy scales in the hole gas.}
As discussed above, the analysis of the PL in magnetic field
allowed us to determine different characteristic energies of the
completely or partially polarized hole gas.

First of all, the reliable determination of the hole density
allows us to calculate the Fermi energy at zero field in the
single particle approach:

$$
E_{\mathrm{F}}=\frac{\pi\hbar^{2}p}{m_{hh}}.
$$

In the fully polarized gas, it is 2\emph{E}$_{\mathrm{F}}$. As
mentioned in part III, depending on the description of the valence
band, the heavy-hole effective mass might vary from 0.17~$m_{0}$
up to 0.25~$m_{0}$. The range of possible values of the Fermi
energy is marked by the gray area in Fig.12c.

The splitting of the double line (D$_{hi}$ - D$_{low}$) gives the
energy of the excited state of the hole gas, which is the final
state in the transition related to the lower component D$_{low}$.
This excitation involves no spin flip, i.e., all carriers are in
the majority spin subband. This energy increases with the carrier
density, as shown in Fig.12c. In a single particle band-to-band
description, an electron at \emph{k}=0 may recombine with any
hole with wavevector between \emph{k}=0 and \emph{k
}=\emph{k}$_{\mathrm{F}}$, particularly if the rule of
conservation of the wavevector in the optical transition is
relaxed by disorder. In a non polarized gas, the intensity is
enhanced at the low energy edge due to a larger recombination
rate, and at the high energy edge due to the so-called Fermi edge
singularity. Then the width of the line is equal to the Fermi
energy of the hole gas. One could imagine to have similar effects
in a polarized gas (with a width equal to
2\emph{E}$_{\mathrm{F}}$ for a fully polarized gas).

It is clear in Fig.12c that at low carrier density the splitting
of the double line is larger than the Fermi energy of the hole
gas, and it appears to deviate from linearity. Recombination of
electrons with a finite wavevector (due to slow relaxation) would
add to that, the kinetic energy of electrons up to
\emph{k}$_{\mathrm{F}}$. This is however unlikely since the double
line appears as a whole at the crossing, with a uniform lifetime
measured in time resolved PL. This suggests that the initial state
is unique, one possibility being the triplet state of the exciton,
which is thought to have in some cases a small but finite binding
energy with respect to the free carrier continuum. The energy in
the final state should be discussed in terms of excitations of the
hole gas (plasmon, combination of single particle excitations,
many body excitations...) \cite{Jusserand2001} with a total
wavevector equal to \emph{k}$_{\mathrm{F}}$. At large carrier
density, the (D$_{hi}$ - D$_{low}$) splitting tends to match the
Fermi energy, as expected for simple band-to-band transitions.

It is thus difficult to precise the nature of the initial state of
the double line, but for the fact that it involves holes sitting
all on the majority spin subband. At very low carrier density, we
have shown that the neutral exciton is most probably involved. It
is not clear whether a triplet state of the charged exciton could
be stabilized by the Zeeman energy. Up to now, triplet states have
been described in non-magnetic QWs at field values large enough to
alter the orbital motion of the carriers
\cite{Shields95a,Crooker00,Homburg2000,Riva2001,Sanvitto2002},
which is not the case here. At large carrier densities, these
bound states will become more and more difficult to distinguish
from an uncorrelated electron-hole pair, but for the fact that in
the absence of interactions, the high energy component D$_{hi}$
should vanish. We have found that the relative intensity of the
two components of the double line changes from sample to sample.
In the case of a band-to-band transition in the presence of a
non-polarized gas, the high energy component is expected to be
enhanced by disorder and by the formation of the Fermi edge
singularity at low carrier density (which eventually evolves into
the charged exciton transition). The situation is different in the
present case of a polarized gas. We clearly and systematically
observe that the intensity of this high energy component increases
when the carrier density is decreased in one sample (by barrier
illumination or by using a \emph{pin} structure). But the intense
line which subsists alone at very small carrier density is not the
singlet state of the charged exciton, but the neutral exciton (or
possibly a triplet state of the charged exciton, as discussed
above).

We have no indication of a relationship with disorder (which could
be induced by alloy fluctuations at larger Mn content or by the
electrostatic disorder known to be present in the \emph{pin}
samples). In such disordered samples, we generally find that the
charged exciton persists at larger values of the spin splitting.
This might be an indication that in those samples, disorder more
severely affects the bound states, likely to be easier to
localize. This is in particular the case of samples with a larger
Mn content, which display carrier induced ferromagnetic
interaction \cite{Haury97}. In such samples, we have indeed
observed the jump from the charged exciton PL to the double line:
it could be induced not only by the giant Zeeman splitting under
applied magnetic field, as in the present study, but also by the
splitting due to the spontaneous magnetization.

Another characteristic energy is the valence band Zeeman splitting
for which the hole gas is completely polarized. It is determined
by the analysis of the evolution of the position of the PL and
absorption lines. At this particular field, and in a single
particle approach, the Zeeman splitting of the valence band is
equal to the Fermi energy. The comparison between the Zeeman
splitting (which is well known in (Cd,Mn)Te), and the Fermi
energy, is given in Fig.12c. We find that the single particle
Fermi energy is over two times larger than the valence band
splitting at complete polarization. Such an enhancement of the
susceptibility of a carrier gas is known to be the result of many
body interactions \cite{ Altshuler85, Zhu2003, Junwirgh01,
Dietl97}. It is usually calculated for an electron gas. Our system
gives a very unique and direct insight to those effects in a hole
gas.

With this in mind, it is worth to come back to the discussion of
the final state involved in the $\sigma^+$ transition from the
charged exciton. This state was described in terms of single
particle excitations in section IV-C. However, one should note
that this description holds only if one applies to the energy of
these excitations the same reduction factor as for the Zeeman
splitting needed to get full polarization (i.e., the inverse of
the enhancement of spin susceptibility).

\subsection{Spectroscopy and energies related to the charged exciton.}
The dissociation energy of the charged exciton can be measured as
the charged-exciton / neutral-exciton splitting at very low
carrier density. At moderate carrier densities, it was found that
the distance between the charged exciton and the neutral exciton
varies with the carrier concentration. This variation has been
studied in absorption experiments for both X$^{+}$ and X$^{-}$
\cite{Kossacki99,Huard2000} in CdTe based QWs. It is linear with
the density of carriers in the spin subband promoting the
formation of the charged exciton. Similar results have been found
for GaAs \cite{Cole98} and ZnSe-based \cite{Astakhov02b} QWs.

\begin{figure}[h]
\includegraphics[width=60mm]{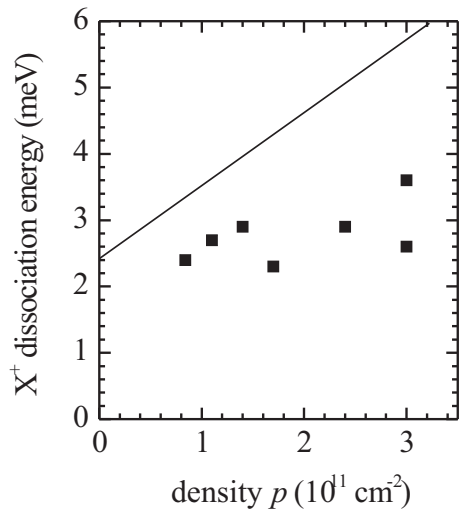}
\caption[]{Dissociation energy of the charged exciton, determined
from the level crossing (squares), versus the carrier density;
the solid line shows the X-X$^{+}$ splitting determined from
absorption spectra in reference~\cite{Kossacki99}.} \label{fig13}
\end{figure}

In this work we have shown, that the jump from the single X$^{+}$
line and the double D$_{hi}$ - D$_{low}$ lines, observed in
$\sigma^+$ polarisation, measures the binding energy of the
charged exciton (The Zeeman splitting of the valence band at this
field gives directly the zero-field splitting of the levels).
This dissociation energy is plotted versus the carrier density in
Fig. 13. No significant variation was observed, the splitting
staying between 2 and 3~meV for all measured samples.

The apparent discrepancy between the line splitting in absorption
spectra and this result of an analysis of the level crossing leads
to the conclusion that in absorption, the system is left in an
excited state which is at higher energy than the energy of the
neutral exciton alone. This is in agreement with theoretical works
\cite{Suris2001,Esser2001} which demonstrate that the increase of
the line splitting measured on the spectra is due to a transfer of
oscillator strength to a high-energy tail of the exciton line,
which contains exciton-carrier scattering states (or un-bound
states of the trion). Note that in the present study, we have an
un-ambiguous determination of the binding energy of the charged
exciton, but no real determination of the excited state, with
respect to which this binding energy is measured. We simply know
that this excited state contains holes all in the majority spin
subband, and is close to the neutral exciton at low carrier
density and to uncorrelated electron-hole pair at high carrier
density.

\section{Summary.}
We have studied a series of Cd$_{1-x}$Mn$_{x}$Te QWs, with
\emph{x} below 0.01, containing a hole gas of density up to
5$\times$10$^{11}$~cm$^{-2}$. In all samples, we observe either a
dichroic line (one line in each of the $\sigma^+$/$\sigma^-$
polarizations, with a splitting related to the giant Zeeman
effect) at low values of the hole spin splitting, and  a double
line in $\sigma^+$ polarization only, at large spin splitting. In
the first case, the spectra continuously emerge from the X$^{+}$
charged exciton lines as the carrier density increases, and keep
several characteristic features of the trion PL. In the other
case, they resemble spectra due to band-to-band PL transitions.

The jump from the trion-like spectra to the double-line is due to
a crossing of the initial levels of the transitions. The charged
exciton contains two holes arranged in a singlet state, which
costs an energy equal to the hole spin splitting. When the hole
spin is larger than the dissociation energy of the charged
exciton, the charged exciton is destabilized in favor of a state
where all holes are in the majority spin subband. We note that
such large values of the spin splitting at low field can be
achieved only in diluted magnetic semiconductors with a hole gas:
In the case of an electron gas, at least four times larger Mn
contents would have to be used. We find that the dissociation
energy of the charged exciton coincides with the X / X$^{+}$
spectral splitting only at low carrier density. At moderate
carrier density, the spectral splitting increases with the carrier
density (in agreement with the results of theoretical studies
which consider excitations accompanying the creation of the
excitons), while the dissociation energy stays constant.

At very low spin splitting, the PL line and the absorption line do
not coincide in energy ("excitonic Moss-Burstein shift"). This we
ascribe to the fact that the position of the absorption line
involves the creation of  excitations of the hole gas in the
presence of a charged exciton at Fermi wavevector, and that the
position of the PL line involves excitations of the partially
polarized hole gas resulting from the recombination of a charged
exciton at the center of the Brillouin zone. Hence a new picture
emerges from the present study for the neutral and charged
excitons at moderate carrier density, with a constant binding
energy of the trion, and various excitations of the hole gas
involved in the transitions. The presence of this "excitonic
Moss-Burstein shift" can be used to determine the carrier density
(once a calibration has been performed), but also to determine the
spin splitting which provokes the complete polarization of the
hole gas. We experimentally find that, in the range of hole
densities explored (from 0.5$\times$10$^{11}$ to
4$\times$10$^{11}$~cm$^{-}2$), it increases with the carrier
density, but remains smaller than the (doubled) Fermi energy. This
enhancement of the spin susceptibility is by a factor at least
larger than 2. However, in agreement with Kohn's theorem (which
might not apply in the valence band...), the observed
spectroscopic splitting is not changed.

The double-line observed in $\sigma^+$ polarization at a
large-enough spin splitting displays several features of a
band-to-band transition. The high-energy component D$_{hi}$ has
the ground state as the final state, and it is associated to a
strong LO-phonon replica: in a band-to-band picture, this involves
an electron relaxed at \emph{k}=0 in the conduction band
recombining with a hole at Fermi wavevector in the valence band,
so that the final state is the ground state. This transition,
indirect in the reciprocal space, can be made slightly allowed by
disorder and carrier-carrier interactions (Fermi edge
singularity), but also by the creation of a phonon with a
wavevector equal to \emph{k}$_{\mathrm{F}}$. The high-energy
component D$_{hi}$ leaves some excitation in the carrier gas: in
the band-to-band description, this final state corresponds to the
excitation of a hole from \emph{k}=0 to \emph{k}$_{F}$, the
single-particle energy being equal to 2\emph{E}$_{\mathrm{F}}$ in
our notation (i.e., the Fermi energy of the fully polarized hole
gas). We note that this line is observed close to the energy
extrapolated at zero-field for the transitions between Landau
levels, and also that a weak but distinct absorption line is seen
in $\sigma^+$ polarization at complete hole polarization, with a
shift with respect to the present lower component of the PL
doublet nearly equal to the calculated Moss-Burstein shift. Hence
the absorption line, the lower component of the double PL line,
and the absorption lines at integer filling factors, can be
qualitatively understood in terms of electron-holes pairs.

However new features appear when a more quantitative description
is attempted. The main discrepancy is the splitting between the
two components of the double line. It is definitely larger than
the Fermi energy (doubled since at full polarization) expected for
the single particle excitation promoting a hole from from
\emph{k}=0 to \emph{k}$_{F}$. This is particularly evident at
intermediate carrier density while at the higher densities
achieved, and considering the larger fluctuation of our data, it
approaches the expected 2\emph{E}$_{\mathrm{F}}$ value. As we have
shown that the dynamics of the double line implies that a single
initial state is involved, this energy has to be accounted for in
terms of excitations of the hole gas (single particle excitations
or combinations of them, or plasmons, taking into account the
effect of carrier-carrier interactions). The previous discussion
of this double line as being close to a band-to-band transition
suggests that the final state has a total wavevector equal to
\emph{k}$_{F}$.

\begin{acknowledgments}
We wish to thank Ronald Cox, Vincent Huard and Kuntheak Kheng for
allowing us to use their experimental set-up when installed at the
Grenoble High Field Laboratory, and for numerous interesting
discussions. This work is partially supported by the Polish-French
Collaboration Program Polonium, and by the Polish State Committee
for Scientific Research (KBN grant 5 P03B 023 20). We thank also
the Swiss national research foundation and the Federal Office for
Education and Science for partial support.

We wish to dedicate this paper to the memory of Andr\'e Wasiela.
\end{acknowledgments}

\begin{table*}
\caption{\label{Table1} Characteristics of selected samples.
Samples "N" are doped using nitrogen acceptors on both sides of
the QW, samples "S" are doped from surface states, and samples "D"
are pin diodes; numbers were chosen to approximately match the Mn
content}
\begin{ruledtabular}
\begin{tabular}{c|c|c|c|c|c|c}
sample  & QW thickness & QW Mn content  & max hole density  & structure &  growth name &   figure \\
&   (nm)  &  ($\%$)  & (10$^{11}$ cm$^{-2}$) &   &     &   \\
\hline
N0 & 8 & none & ~0.3 & N-doped & M751 & 1 \\
N2 & 8 & 0.18 & 3.1 & N-doped & M968 & 3, 4a, 9a \\
N3 & 8 & 0.37 & 4.2 & N-doped & M1038 &  5, 6, 12 \\
N4 & 8 & 0.40 & 3.2 & N-doped & M1305 &  7, 8, 10, 12, 13 \\
N5 & 8 & 0.53 & 3.8 & N-doped & M1132 & 2, 4b, 9b, 12, 13 \\
N4b & 8 & 0.4 & 5.2 & N-doped & M1131 &  12 \\
S7 & 10 & 0.7 & 2.8 & surface doped & M1269 &  11,12,13 \\
S8 & 10 & 0.7 & 1.5 & surface doped & M1290 &  12, 13 \\
D3 & 8 & 0.3 & 2.8 & pin diode & M1329 & 12, 13 \\
D50 & 8 & 5 & 1.8 & pin diode & M1346 &  12\\
\end{tabular}
\end{ruledtabular}
\end{table*}

\end{document}